\documentclass[noshowpacs,amsmath,
twocolumn,
superscriptaddress,
8pt,
aps,prb]{revtex4-1}
\usepackage{setspace}
\usepackage{amsmath}
\usepackage{graphicx}
\usepackage[nearskip,margin = 0pt]{subfig}

\usepackage{verbatim}
\usepackage{amsfonts}
\usepackage{amssymb}
\usepackage{epstopdf} 
\usepackage{xcolor}
\DeclareGraphicsExtensions{.pdf,.eps,.png,.jpg,.mps} 

\begin{document}

\title{Stokes Solitons in Optical Microcavities}

\author{Qi-Fan Yang$^{\ast}$, Xu Yi$^{\ast}$, Ki Youl Yang and Kerry Vahala$^{\dagger}$\\
T. J. Watson Laboratory of Applied Physics, California Institute of Technology, Pasadena, California 91125, USA.\\
$^{\ast}$These authors contributed equally to this work.\\
$^{\dagger}$Corresponding author: vahala@caltech.edu}
\maketitle

\noindent {\bf Solitons are wavepackets that resist dispersion through a self-induced potential well. They are studied in many fields, but are especially well known in optics on account of the relative ease of their formation and control in optical fiber waveguides \cite{hasegawa1973transmission,mollenauer1980experimental}. Besides their many interesting properties, solitons are important to optical continuum generation \cite{dudley2006supercontinuum}, in mode-locked lasers \cite{haus2000mode, cundiff2005soliton} and have been considered as a natural way to convey data over great distances \cite{haus1996solitons}. Recently, solitons have been realized in microcavities \cite{herr2014temporal} thereby bringing the power of microfabrication methods to future applications. This work reports a soliton not previously observed in optical systems, the Stokes soliton. The Stokes soliton forms and regenerates by optimizing its Raman interaction in space and time within an optical-potential well shared with another soliton. The Stokes and the initial soliton belong to distinct transverse mode families and benefit from a form of soliton trapping that is new to microcavities and soliton lasers in general. The discovery of a new optical soliton can impact work in other areas of photonics including nonlinear optics and spectroscopy.}

\begin{figure*}
    \begin{centering}
  \includegraphics[width=17.5 cm]{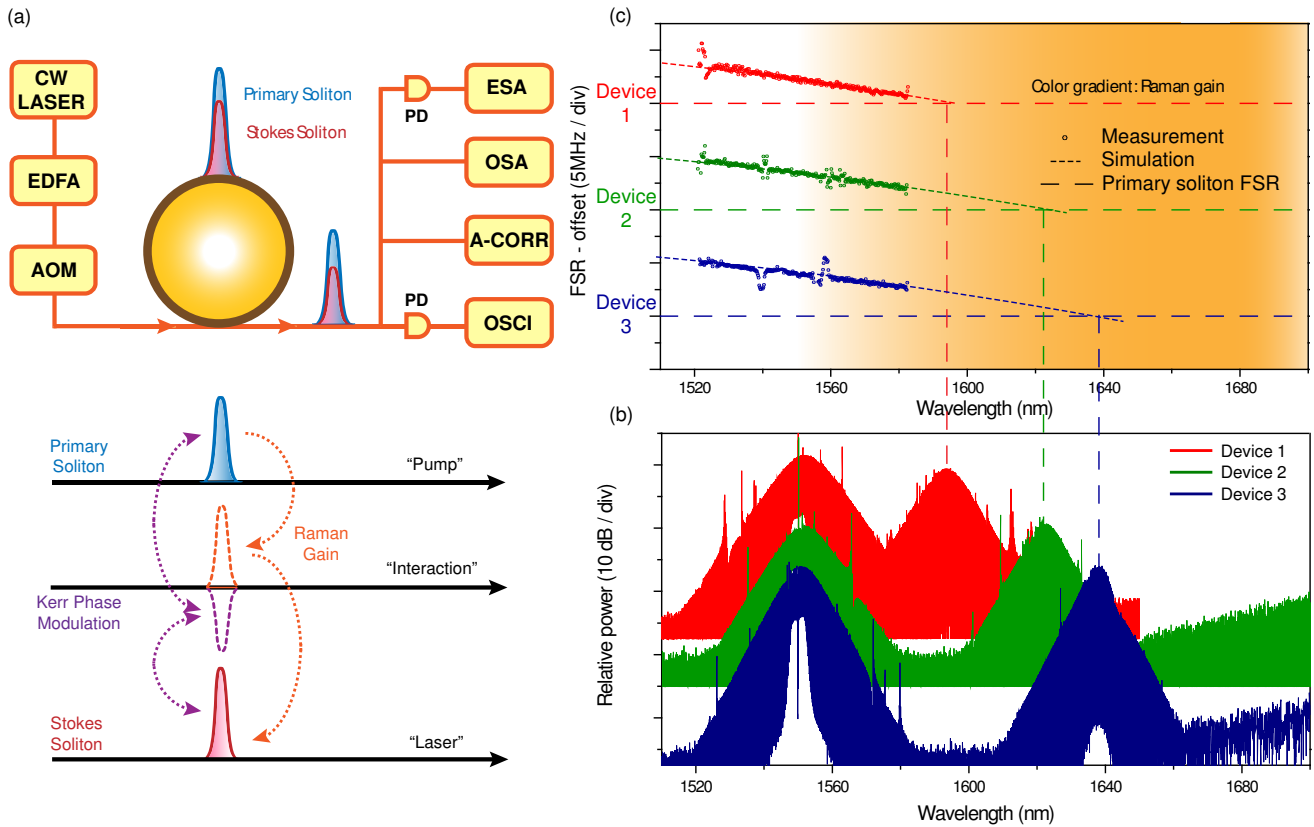}
  \captionsetup{singlelinecheck=no, justification = RaggedRight}
  \caption{\textbf{Experimental setup and description of Stokes soliton generation, primary and Stokes soliton spectra in three microcavity devices, and Stokes soliton mode family FSR dispersion measurement. }{\bf a)} The Stokes soliton (red) maximizes Raman gain by overlapping in time and space with the primary soliton (blue). It is also trapped by the Kerr-induced effective optical well created by the primary soliton. The microcavity (shown as a ring) is pumped with a tunable, continous-wave (CW) fiber laser amplified by an erbium-doped fiber amplifier (EDFA). An acousto-optic modulator (AOM) is used to control the pump power. The output soliton power is detected with a photo diode (PD) and monitored on an oscilloscope (OSCI). Wavelength division multiplexers (not shown) split the 1550 nm band primary soliton and 1600 nm band Stokes soliton so that their powers can be monitored separately on the oscilloscope. An optical spectrum analyzer (OSA), auto-correlator (A-CORR) and electrical spectrum analyzer (ESA) also monitor the output. {\bf b)} Primary and Stokes soliton spectra observed in three devices. {\bf c)} Free spectral range (FSR) versus wavelength measured for mode families associated with the Stokes soliton in three microcavity devices. Extrapolation of the measured results is done using mode simulation (dashed lines).  The FSR near the spectral center of the primary soliton in each device is shown as a dashed horizontal line. The spectral location of the Stokes soliton in (b) closely matches the predicted FSR matching wavelength. The background coloration gives the approximate wavelength range of the Raman gain spectrum.}
  \label{fig1}
    \end{centering}
\end{figure*}

\smallskip

\noindent {\bf Introduction.} Solitons result from a balance of wave dispersion with a non-linearity. In optics, temporal solitons are readily formed in optical fiber waveguides \cite{hasegawa1973transmission,mollenauer1980experimental,haus1996solitons,dudley2006supercontinuum} and laser resonators \cite{haus2000mode,cundiff2005soliton} and have recently been observed in dielectric microcavties \cite{herr2014temporal}. In each of these cases nonlinear compensation of group velocity dispersion is provided by the Kerr effect (nonlinear refractive index).  Besides the Kerr nonlinearity, a secondary effect associated with soliton propagation is the so-called soliton self-frequency shift caused by Raman interactions, which induce a continuously increasing red-shift with propagation in a waveguide \cite{dudley2006supercontinuum} or a fixed shift of the soliton spectrum in cavities \cite{karpov2015raman}. More generally, the Raman interaction can produce optical amplfication and laser action of waves red-shifted relative to a strong pumping wave \cite{islam2002raman}. This work reports a new Raman-related effect, soliton generation through time and space varying Raman amplification created by the presence of a first temporal soliton. Because the new soliton is spectrally red-shifted relative to the initial soliton we call it a Stokes soliton. It is observed in a silica microcavity and obeys a threshold condition resulting from an optimal balancing of Raman gain with cavity loss when the soliton pulses overlap in space and time. Also, the repetition frequency of both the initial and the Stokes soliton are locked by the Kerr nonlinearity. 

In this work the cavity will be a circular-shaped whispering gallery microcavity and the first temporal soliton will be referred to as the primary soliton. The primary soliton here is a dissipative Kerr cavity soliton \cite{herr2014temporal}, however, other solitons would also suffice. Consistent with its formation, the primary soliton creates a spatially varying refractive index via the Kerr nonlinearity that serves as an effective potential well, traveling with the soliton and and counteracting optical dispersion. Moreover, on account of the Raman interaction, the primary soliton creates local Raman amplification that also propagates with the primary soliton. The primary soliton is composed of many longitudinal modes belonging to one of the transverse mode families of the cavity. $\Delta \nu_P$ will be the longitudinal mode separation or free spectral range for longitundinal modes near the spectral center of the primary soliton. $\Delta \nu_P$ also gives the approximate round trip rate of the primary soliton around the cavity ($T_{RT} = \Delta \nu_P^{-1}$ is the round trip time). 

Consider another transverse mode family besides the one that forms the primary soliton. Suppose that some group of longitudinal modes in this family satisfies two conditions: (1) they lie within the Raman gain spectrum created by the primary soliton; (2) they feature a free spectral range (FSR) that is close in value to that of the primary soliton ($\Delta \nu_P$). Any noise in these longitudinal modes will be amplified by Raman gain provided by the primary soliton.  If the round trip amplification of a resulting waveform formed by a superposition of these modes is sufficient to overcome round trip optical loss, then oscillation threshold is possible. The threshold will be lowest (Raman gain maximal) if the modes of the second family phase lock to form a pulse overlapping in both space and time with the primary soliton. This overlap is possible since the round trip time of the primary soliton and the new pulse are closely matched, i.e., condition (2) above is satisfied. Also, the potential well created by the primary soliton can be shared with the new optical pulse. This latter nonlinear coupling of the primary soliton with the new, Stokes soliton pulse results from Kerr-mediated cross-phase modulation and further locks the round trip rates of the two solitons (i.e., their soliton pulse repetition frequencies are locked). As an aside, a third condition on the new mode family that forms the Stokes soliton is that it feature spatial overlap with the spatial intensity distribution of the primary soliton transverse mode family. A conceptual schematic of the process is presented in figure 1a. 

The generation of a fundamental soliton by another fundamental soliton in this way is new and also represents a form of mode locking of a soliton laser. It differs from mechanisms like soliton fission which also result in the creation of one of more fundamental solitons \cite{dudley2006supercontinuum}. Specifically, soliton fission involves a higher order soliton breaking into mutliple fundamental solitons, nor is it a regenerative process. Also, whereas Raman self-frequency shifting in solitons is well known in optical fibers \cite{mitschke1986discovery,gordon1986theory} and has been recently observed in optical microcavities  \cite{yi2015soliton,karpov2015raman}, the Raman mediated formation and regeneration of a new soliton by an existing soliton is new. Finally, concerning the trapping phenomena that accompanies the Stokes soliton formation, the trapping of temporal solitons belonging to distinct polarization states was proposed in the late 1980s \cite{menyuk1987stability} and was observed in optical fiber \cite{islam1989soliton, cundiff1999observation} and later in fiber lasers \cite{collings2000polarization}. However, trapping of temporal solitons belonging to distinct transverse mode families, as oberved here, was proposed even earlier\cite{hasegawa1980self,crosignani1981soliton}, but has only recently been observed in graded-index fiber waveguides \cite{renninger2013optical} and not so far in a laser or a cavity. The observation, measurement and modeling of Stokes solitons is now described. 

\smallskip

\begin{figure*}
    \begin{centering}
  \includegraphics[width=17.5 cm]{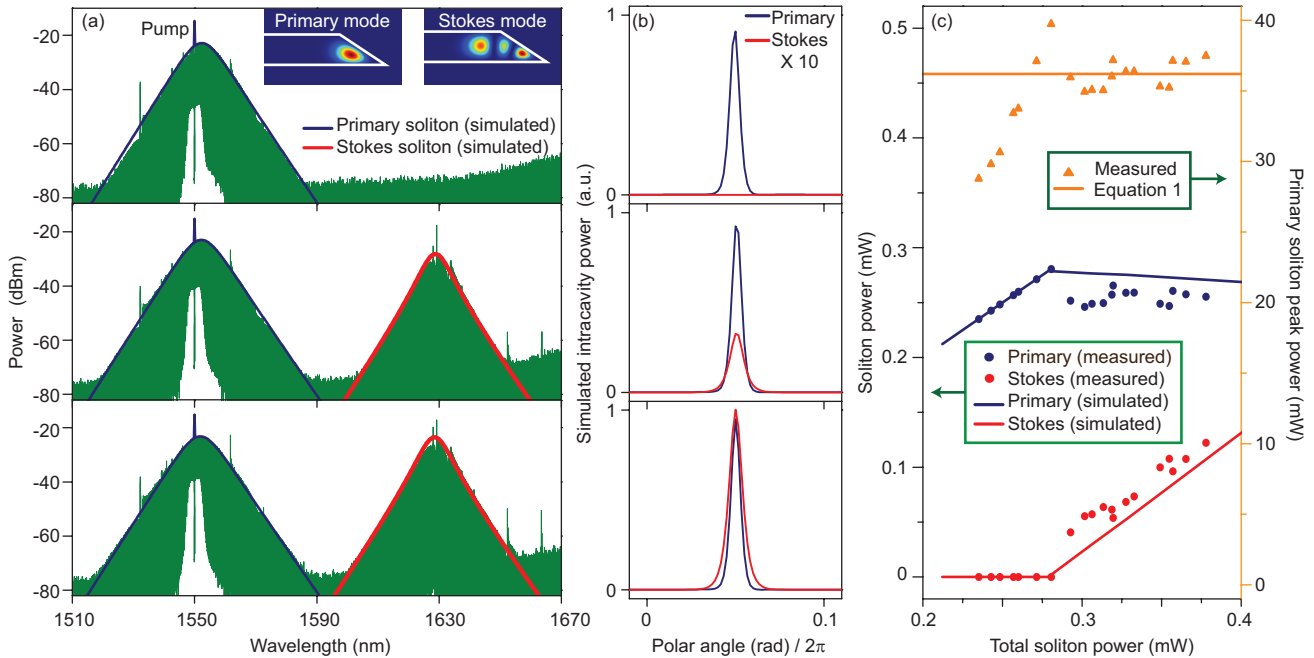}
  \captionsetup{singlelinecheck=no, justification = RaggedRight}
  \caption{\textbf{Stokes soliton spectra, power and threshold measurements.} {\bf a} Soliton spectra are plotted at three primary soliton powers (one below the Stokes soliton threshold). The insets show the spatial mode families associated with the primary and Stokes solitons. The blue and red curves are simulations using the coupled Lugiato Lefever equations.  {\bf b} Simulated primary and Stokes soliton pulses corresponding to the spectral simulations in (a) plotted versus their angular location in the cavity. {\bf c} Measurement and theory of Stokes (red) and primary (blue) soliton power versus total soliton power. The primary soliton peak power (orange) versus total power is also plotted to show threshold clamping at the onset of Stokes soliton oscillation. The theoretical threshold peak power from eqn. (1) is also shown for comparison.} 
  \label{fig2}
    \end{centering}
\end{figure*}

\begin{figure}
    \begin{centering}
  \includegraphics[width=8.5 cm]{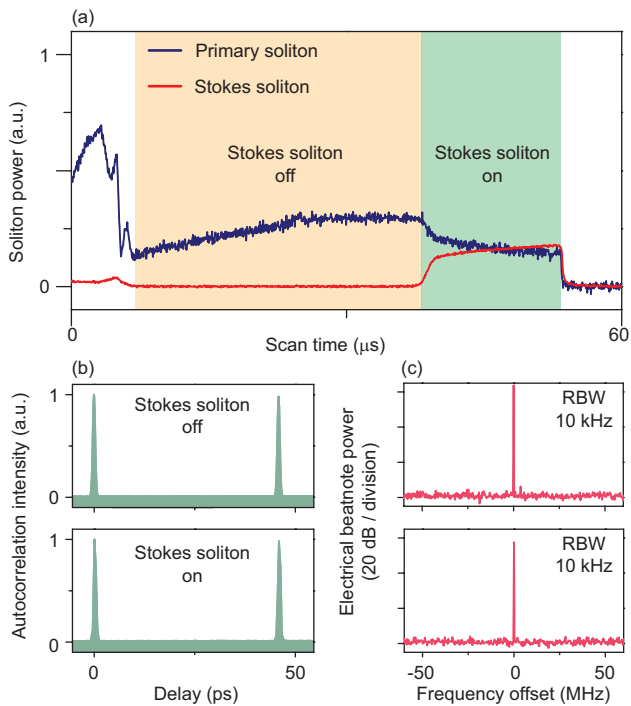}
  \captionsetup{singlelinecheck=no, justification = RaggedRight}
  \caption{\textbf{Scanned soliton power, autocorrelation measurements and detection of primary and Stokes soliton microwave repetition rate. } \textbf{a,} Primary and Stokes soliton power scanned versus time. Regions to the far left in the scan correspond to pumping conditions that do not form a primary soliton. Regions to the far right in the scan are beyond the existence limit of the primary soliton.  \textbf{b,} Autocorrelation measurements of the primary soliton both below and above Stokes soliton oscillation threshold. \textbf{c,} Electrical spectrum of the detected primary soliton pulse stream (upper) and the combined primary and Stokes pulse streams (lower). The zero on the frequency scale corresponds to 22 GHz.}
  \label{fig3}
    \end{centering}
\end{figure}

\noindent {\bf Observation of Stokes Solitons.} The experimental setup is shown in figure 1a.  The microcavities are about 3mm in diameter, fabricated from silica on silicon and have an unloaded optical Q factor of 400 million\cite{lee2012chemically}.  The cavity is pumped by a continuous-wave fiber laser to induce a primary soliton which is a dissipative Kerr cavity (DKC) soliton with a repetition frequency of 22GHz and pulse width that could be controlled to lie within the range of 100-200 fs.  The requirements for DKC soliton generation include a mode family that features anomolous dispersion. Other requirements as well as control of DKC soliton properties are described elsewhere\cite{herr2014temporal,yi2015soliton}. DKC soliton spectra observed in three different microcavity devices are shown in figure 1b (red, green and blue spectra centered near 1550nm). As the primary soliton power is increased, the Stokes soliton threshold is achieved and its spectrum is observed at longer wavelengths relative to the primary soliton (figure 1b). Note that condition 1 (Raman gain) is satisified as the new spectra lie within the Raman gain spectrum created by the primary soliton (see shaded area in figure 2c). 

The dispersion in the FSR of the Stokes soliton mode family is measured in figure 1c to confirm condition 2 (matching FSR of primary and Stokes solitons).  To measure the mode families a tunable external cavity diode laser (ECDL) scans the spectral locations of optical resonances from 1520 nm to 1580 nm. The resonances appear as minima in the optical power transmitted past the microcavity, and the location of these resonances is calibrated using a fiber-based Mach Zehnder interferometer.  The resulting data provide the dispersion in the FSR of cavity modes versus the wavelength and readily enable the identification transverse mode families. The results of measurements on the three devices are summarized in figure 1c. The FSR at the spectral center of the primary soliton is plotted as a horizontal dashed line (red, green, blue according to device) while the FSR of a neighboring mode family is plotted versus wavelength. Even though the range of the measurement is not sufficient to extend into the wavelength band of the observed Stokes soliton, an extrapolation of the data is performed using a modal dispersion simulation\cite{yi2015soliton}.  As can be seen, the wavelength where primary and Stokes solitons FSR are equal is close to the spectral maximum of the corresponding Stokes solitons in figure 1b. As a final comment, primary and Stokes solitons are observed to have the same polarization.

The thresholding nature of Stokes soliton formation is observed in figure 2a. The spectra show the primary soliton spectra measured at three power levels and the corresponding Stokes soliton spectra. One spectral trace is measured for a power level below the Stokes soliton threshold. The spectral width and the power in all spectral lines of the Stokes solitons grow with pumping above threshold. This behavior is distinct from that of DKC solitons, where increased soliton power is accompanied only by increasing soliton spectral width (i.e., maximum power per frequency line is fixed). Simulations of the spectra and the corresponding pulses are also provided in figure 2a and figure 2b. These use the coupled Lugiato-Lefever equations (Methods Section).  In figure 2c power data are provided showing the primary soliton peak and average power as well as the Stokes soliton power plotted versus the total soliton power. Simulations are provided for comparison with the power data. As an aside, the Stokes soliton power can exceed the power of the primary soliton. This was observed in one device and is a consequence of the threshold condition.

The threshold-induced gain-clamping condition imposed on the primary soliton (which functions as a pump for the Stokes soliton) is readily observable in figure 2c. As shown in the Methods Section, this threshold condition clamps the peak power of the primary soliton and is given by the following expression,
\begin{equation}
P_{th}=\frac{\kappa_p^{ext}\kappa_s }{2 R}(1+\frac{1}{2 \gamma})
\end{equation}
where $P_{th}$ is the threshold peak output power of the primary soliton and other parameters are defined in the Methods Section. Eqn. (1) is derived assuming a weak Stokes soliton power relative to the primary soliton which is an excellent assumption near threshold. Eqn. (1) is plotted for comparison to the peak power data using the same parameters used in the figure 2c simulation plots. 

Another way to measure the relationship between power in the primary and Stokes solitons is to monitor the average power in their respective pulse trains while the power in primary soliton is being scanned. Because the primary soliton is a DKC soliton its power is varied by scanning the detuning of the pump laser relative to the pumping mode of the primary soliton\cite{yi2015soliton}. The onset of the DKC soliton oscillation occurs when the scanned pump is red-detuned relative to the resonance, and its power increases as the pump continues to detune further to the red. Utlimately, the existence detuning limit is reached and the primary soliton shuts off \cite{yi2015soliton}. In figure 3a a temporal scan shows the primary soliton increasing in power and the appearance of the Stokes soliton towards the end of the scan. Because it is the peak power of the primary soliton that is clamped above the Stokes soliton threshold, the {\it total} primary soliton power actually decreases once the Stokes turns on. This happens because the primary soliton pulse width changes during the scan. 

Autcorrelation measurements of the primary soliton both with and without the Stokes soliton are presented in figure 3b. Similar measurements were not possible for the Stokes soliton on account of the wavelength limitations of the autocorrelation system. However, broad optical band detection of the soliton repetition rate was possible for both the primary and Stokes solitons pulse streams. Figure 3c shows the corresponding electrical spectrum analyzer signal for detection of the primary soliton signal and for simultaneous detection of the both the primary and Stokes soliton signals. The frequency of the beat note is identical in both cases confirming that the corresponding soliton round-trip rates are locked.

\smallskip

\noindent {\bf Summary.} The Stokes soliton is only the second type of soliton to be observed in microcavities (beyond dissipative Kerr solitons \cite{herr2014temporal}) and also represents the first time soliton trapping has been observed in any microcavity. It also represents the first observation of trapping by solitons in different transverse modes in a laser. From a practical viewpoint, the pilot and primary solitons overlap in space and time, and have a frequency separation that can be engineered to fall within the mid IR range. As a result, this soliton system is potentially interesting for mid IR generation by way of difference frequency generation. Not all devices are observed to produce Stokes solitons. However, dispersion engineering techniques are being advanced \cite{yang2015broadband} and should enable control of both observation of the Stokes soliton as well as its placement in the optical spectrum. Indeed, the spectral placement of Stokes solitons in figure 1b is largely the result of microcavity diameter control to shift the FSR crossing point (condition 2). The specific implementation described here uses a compact microresonator on a silicon wafer which also suggests that monolithic integration will be possible. In an appropriately phase-matched multimode waveguide (optical fiber or monolithic) it should also be possible to observe non-cavity-based Stokes solitons. 

\smallskip

\noindent {\bf Methods}

\begin{footnotesize}

\noindent {\bf Coupled Lugiato Lefever equations.} A pair of coupled equations describing the intracavity slowly-varying field amplitudes for the primary and Stokes soliton system can be found from the Lugiato-Lefever equation (LL equation) \cite{lugiato1987spatial, matsko2011mode, chembo2013spatiotemporal, herr2014temporal} augmented by Raman terms \cite{agrawal2007nonlinear,bao2014nonlinear}. By including cross-phase modulation and Raman interaction terms the following system of equations results:
\begin{equation}
\begin{split}
\frac{\partial E_p}{\partial t}=&i \frac{D_{2p}}{2}\frac{\partial^2 E_p}{\partial \phi^2}+i g_p\left|E_p\right|^2 E_p+i G_p\left|E_s\right|^2 E_p\\&-i g_p D_{1p}\tau_R E_p\frac{\partial(\left| E_p\right|^2+A_{pp}\left|E_s\right|^2/A_{ps})}{\partial\phi}\\&-(\frac{\kappa_p}{2}+i\Delta\omega_p) E_p-\frac{\omega_p}{\omega_s}R|E_s|^2 E_p+\sqrt{\kappa^{ext}_p P_{in}},
\end{split}
\end{equation}
\begin{equation}
\begin{split}
\frac{\partial E_s}{\partial t}=&-\delta\frac{\partial E_s}{\partial\phi}+i \frac{D_{2s}}{2}\frac{\partial^2 E_s}{\partial \phi^2}+i g_s \left| E_s\right|^2 E_s+i G_s\left| E_p\right|^2 E_s\\&-i g_s D_{1p} \tau_R E_s\frac{\partial(\left| E_s\right|^2+A_{ss}\left|E_p\right|^2/A_{ps})}{\partial\phi}\\&-(\frac{\kappa_s}{2}+i\Delta\omega_s) E_s+R|E_p|^2 E_s,
\end{split}
\end{equation}
The slowly varying fields $ E_j $ (subscript $j= (p,s)$ for primary or Stokes soliton) are normalized to optical energy. To second order in the mode number the frequency of mode number $\mu$ in soliton $j=(p,s)$ is given by the Taylor expansion $\omega_{\mu j}=\omega_{0j}+D_{1j}\mu+\frac{1}{2}D_{2j}\mu^2$ where $\omega_{0j}$ is the frequency of mode $\mu = 0$, while $D_{1j}$ and $D_{2j}$  are the FSR and the second-order dispersion at $\mu = 0$. Also,  $\delta=D_{1s}-D_{1p}$ is the $FSR$ difference between primary and Stokes solitons at mode $\mu =0$. $\kappa_j$ is the cavity loss rate and $\Delta\omega_j$ is the detuning of mode zero of the soliton spectrum relative to the cold cavity resonance. For the primary soliton, which is a DKC soliton, the pump field is locked to one of the soliton spectral lines and this ``pump" line is taken as mode zero. $\tau_R$ is the Raman shock time, $\kappa^{ext}_j$ is the external coupling coefficient and $P_{in}$ is the pump power. $g_j$ and $G_j$ are self and cross phase modulation coefficients, defined as,
\begin{equation}
g_j=\frac{n_2\omega_j D_{1j}}{2n\pi A_{jj}}, \\ \quad
G_j=\frac{(2-f_R)n_2\omega_j D_{1j}}{2n\pi A_{ps}}.
\end{equation}
where the nonlinear mode area $A_{jk}$ is defined as\cite{agrawal2007nonlinear}
\begin{equation}
A_{jk}=\frac{\iint_{-\infty}^{\infty} |u_j(x,y)|^2 dxdy\iint_{-\infty}^{\infty} |u_k(x,y)|^2 dxdy}{\iint_{-\infty}^{\infty} |u_j(x,y)|^2|u_k(x,y)|^2 dxdy},
\end{equation}
where $u_j$ is the transverse distribution of the mode. $f_R=0.18$ is the Raman contribution parameter in silica. $R=cD_{1p} g_{R}(\omega_s,\omega_p)/4n\pi A_{ps} $ where $g_{R}(\omega_s,\omega_p)$ is the Raman gain in silica. For solitons with a few THz bandwidth, other nonlinear effects are negligible (e.g., higher order dispersion, the self-steepening effect and Raman induced refractive index change \cite{agrawal2007nonlinear}). Phase-sensitive, four-wave-mixing terms have been omitted in Eqn. (2) and (3). In principle, these terms could introduce locking of the Stokes and primary soliton fields (in addition to their repetition rates). However, for this to occur the underlying spatial mode families would need to feature mode frequencies that align reasonably well (both in FSR and offset frequency) within the same band. Such conditions do not seem likely even though they might occur accidentally or through dispersion engineering. In this work, one device featured primary and Stokes solitons with overlapping spectra (see device 1 in figure 1b). As expected the mode frequences associated with each soliton did not overlap.

The coupled equations are numerically studied using the split-step Fourier method. Over 600 modes are included in the simulation for each soliton. Parameters are given below. Note that the detuning of the Stokes soliton determines the rotation frame, which can be set to zero during the simulation.

\smallskip

\noindent {\bf Calculation of threshold.} The behavior of the Stokes soliton system can be studied analytically beginning with the coupled LL equations. Near threshold the Stokes soliton is much weaker in power than the primary soliton. In this limit, the cross-phase modulation and cross Raman interaction terms within the primary soliton equation can be neglected, while the self-phase modulation and self-Raman terms are neglected in the Stokes equation. The primary soliton is then governed by the standard uncoupled LL equation, which features an analytical solution of the field amplitude (hyperbolic secant form)\cite{herr2014temporal}. This solution is substituted into the Stokes soliton equation which then takes a Shr$\rm \ddot o$dinger-like form containing a sech$^2$ potential in addition to sech$^2$ Raman gain. The specific solution to these two equations in the Stokes low power limit take the following form:
\begin{equation}
 E_p\sim\mathrm{sech} B\phi,\\
 \quad
 E_s\sim\mathrm{sech}^{\gamma} B\phi,
\end{equation}
where $B=\sqrt{2\Delta\omega_{p}/D_{2p}}$. The Stokes soliton solution is the general solution for a soliton trapped in the index well created by the sech$^2$ primary soliton intensity where the power $\gamma$ satisfies the equation,
\begin{equation}
\gamma ( \gamma +1)=2G_s D_{2p}/g_p D_{2s}.
\end{equation}
Once the peak power of the primary soliton reaches a point that provides sufficient Raman gain to overcome roundtrip loss, the Stokes soliton will begin to oscillate. The threshold condition emerges as the steady state Stokes soliton power balance. This is readily derived from the Stokes soliton equation and takes the form,
\begin{equation}
\int_{0}^{2 \pi} \partial_t |E_s|^2 d\phi=\int_{0}^{2 \pi}d\phi(\kappa_s -2R|E_p|^2)|E_s|^2 = 0 
\end{equation}
By substituting the solutions (eqns. (6)) into eqn. (8), the resulting threshold in primary soliton peak output power is found to be given by eqn. (1).

\smallskip

\noindent {\bf Parameters.} The measured parameters are: $\kappa_p/2\pi=736$ kHz, $\kappa_p^{ext}/2\pi=302$ kHz, $\lambda_p=1550$ nm, $D_{1p}/2\pi=22$ GHz and $D_{2p}/2\pi=16.1$ kHz. Neither the intrinsic or coupling loss of the Stokes mode could be measured in the 1600 nm band. However, the optical loss for the Stokes family could be measured in the 1550 nm band. Accordingly, intrinsic loss of $\kappa_s^o/2\pi=1.11$ MHz was used while the external loss of $\kappa_s^{ext}=1.25$ MHz was slightly tuned to obtain the best fitting. The latter is expected to shift somewhat from 1550 nm to 1600 nm bands due to small changes in phase matching of the microcavity to waveguide coupling. $P_{in}=150$ mW is used throughout the simulation. Calculated parameters (based on mode simulations) are: $D_{2s}/2\pi=21.7$ kHz, $A_{ss}=69.8$ $\mathrm{\mu m^2}$, $A_{pp}=39.7$ $\mathrm{\mu m^2}$, $A_{ps}=120$ $\mathrm{\mu m^2}$, and $\delta=0$ when $\lambda_s=1627$ nm. Other constants are: $n=1.45$, $n_2=2.2\times10^{-20}$ $\mathrm{m^2/W}$, $g_{R}=3.94\times10^{-14}$ $\mathrm{m/W}$, $\tau_R=3.2$ fs. The calculated $\gamma=0.55$.

\end{footnotesize}

\medskip

\noindent {\bf Acknowledgments} The authors thank Steven Cundiff at the University of Michigan for helpful comments on this manuscript. The authors gratefully acknowledge the Defense Advanced Research Projects Agency under the QuASAR program and PULSE programs, NASA, the Kavli Nanoscience Institute and the Institute for Quantum Information and Matter, an NSF Physics Frontiers Center with support of the Gordon and Betty Moore Foundation.

\bibliography{ref}

\end{document}